\documentclass[runningheads]{llncs}
\usepackage{multirow} 
\usepackage{booktabs} 
\usepackage{xcolor}
\usepackage{graphicx}
\usepackage{amsmath} 
\usepackage{caption}
\usepackage{subcaption} 
\usepackage{hyperref}
\usepackage{booktabs} % For better horizontal lines
\usepackage{makecell} % For line breaks in cells
\usepackage{multirow} % For multirow feature
\usepackage{bbding}
\usepackage{adjustbox}
\begin{document}
\title{TeleOR: Real-time Telemedicine System \\for Full-Scene Operating Room}
% \author{Anonymous}
% \institute{Submission \#757}
% \author{Anonymous submission\inst{1} \and

\titlerunning{TeleOR: Real-time Remote Operating Room}
\authorrunning{Y. Wu et al.}

\author{Yixuan Wu\inst{1,2,3}\thefootnote{*} \and % 1{Wu, Yixuan} 
Kaiyuan Hu\inst{4}\thefootnote{*}\and % 2{Hu, Kaiyuan}
Qian Shao\inst{5} \and % 3{Shao, Qian} 
Jintai Chen\inst{6} \inst{(}\Envelope\inst{)}\and \\ % 4{Chen, Jintai}
Danny Z. Chen\inst{7} \and % 5{Chen, Danny}
Jian Wu\inst{1,2}\inst{(}\Envelope\inst{)} } % 6{Wu, Jian}
\institute{School of Public Health, Zhejiang University, China\and
State Key Laboratory of Transvascular Implantation Devices of The Second Affiliated Hospital, Zhejiang University School of Medicine, China\and
Institute of Wenzhou, Zhejiang University, China\and
School of Science and Engineering, The Chinese University of Hong Kong, Shenzhen, China\and
College of Computer Science and Technology, Zhejiang University, China\and
Computer Science Department, University of Illinois at Urbana-Champaign, USA \and
Department of Computer Science and Engineering, University of Notre Dame, USA 
    \email{wyx\_chloe@zju.edu.cn} }
%\email{wujian2000@zju.edu.cn}}

\maketitle              % typeset the header of the contribution
\def\thefootnote{*}\footnotetext{Equal Contributions.}
% \vspace{-2em}
\begin{abstract}
\vspace{-2em}
The advent of telemedicine represents a transformative development in leveraging technology to extend the reach of specialized medical expertise to remote surgeries, a field where the immediacy of expert guidance is paramount. However, the intricate dynamics of Operating Room (OR) scene pose unique challenges for telemedicine, particularly in achieving high-fidelity, real-time scene reconstruction and transmission amidst obstructions and bandwidth limitations. This paper introduces TeleOR, a pioneering system designed to address these challenges through real-time OR scene reconstruction for Tele-intervention. TeleOR distinguishes itself with three innovative approaches: dynamic self-calibration, which leverages inherent scene features for calibration without the need for preset markers, allowing for obstacle avoidance and real-time camera adjustment; selective OR reconstruction, focusing on dynamically changing scene segments to reduce reconstruction complexity; and viewport-adaptive transmission, optimizing data transmission based on real-time client feedback to efficiently deliver high-quality 3D reconstructions within bandwidth constraints. Comprehensive experiments on the 4D-OR surgical scene dataset demostrate the superiority and applicability of TeleOR, illuminating the potential to revolutionize tele-interventions by overcoming the spatial and technical barriers inherent in remote surgical guidance.

\end{abstract}
\vspace{-2em}

\section{Introduction}

The emergence of telemedicine marks a pivotal advancement in utilizing technology to broaden the accessibility of specialized medical expertise~\cite{wu2024aienhancedvirtualrealitymedicine,wootton2001telemedicine,ekeland2010effectiveness}. This shift is crucial for providing timely expert guidance for remote surgery, especially in scenarios where the distance from specialized professionals could delay vital treatment.
Telemedicine encompasses a broad range of disciplines, from social and perceptual to behavioral and technical aspects of remote healthcare delivery. 
In surgical intervention scenarios, the deployment of telemedicine systems necessitates a series of critical steps, including onsite scene reconstruction, transmission, remote rendering, and intervention. 
A significant technical challenge in telemedicine lies in achieving real-time reconstruction and transmission. This necessity is more pronounced in the context of surgery, as it demands a greater level of real-time responsiveness compared to other telemedicine applications, i.e., ward monitoring or remote consultations.

Previous research has explored various methods by employing multiple static RGB-D cameras to capture and enhance the reconstruction of 3D scenes.
The scenarios are often first modeled by separate point clouds from per-camera viewpoints, which are then fused into a common reference frame using intrinsic and extrinsic parameters~\cite{beck2015volumetric,jin2023capturedisplaysurveyvolumetric}. 
To achieve higher fidelity reconstructions, some researchers proposed and refined the use of neural implicit functions~\cite{gerats2023dynamic,kato2018neural,wang2022neural,wolterink2022implicit} and generative models~\cite{he2023dmcvr,kim2022diffusion}. 
However, these methods traditionally relied on designing sophisticated algorithms for robust visual feature extraction~\cite{he2023dmcvr,wu2023gcl,wu2022self} and multiview calibration~\cite{gerats2023dynamic,hu2023fsvvd}, typically conducted in offline settings. 
On the other hand, certain strategies have aimed to simplify these methods to enhance reconstruction efficiency. However, these simplifications tend to compromise the accuracy of the reconstructions, making them only suitable for static scenes~\cite{stotko2019slamcast,weibel2020artemis} rather than dynamic surgical scene. Furthermore, these approaches~\cite{dou2016fusion4d} may restrict the number of remote clients who can access the system, thereby diminishing their applicability for tele-intervention in surgical settings. 

The Operating Room (OR) presents a dynamic and complex scenario, teeming with medical staff, patients, and an array of medical equipment, which poses more challenges in scene understanding scenarios~\cite{ozsoy2023holistic}.  
Unlike traditional static settings where full-scene capturing is achieved by deploying camera arrays to surround clear capture regions, OR often involves obstructions from large devices, hanging monitors, and moving staff, leading to frequent occlusions of scene capture cameras. 
Furthermore, the balance between high-quality reconstruction and efficient real-time transmission underscores the significant challenges in surgical tele-intervention.
To address these challenges, in this work, we propose TeleOR, a real-time full-scene \textbf{O}perating \textbf{R}oom scene reconstruction system for \textbf{Tele}-intervention. 
TeleOR introduces innovative solutions critical for overcoming the complex challenges of telemedicine in surgical scene.
Its critical designs and insights involve:

\noindent\textbf{Dynamic Self-Calibration:} We introduce an innovative dynamic self-calibration approach for multiview cameras in OR scene. This technique leverages the OR scene's inherent features as reference points for calibration, eliminating the reliance on pre-set markers or physical calibration tools. Consequently, it enables cameras to move and self-calibrate in real-time, effectively avoiding obstructions.

\noindent\textbf{Selective OR Reconstruction:} Recognizing that specific sections of the OR scene remain unchanged for extended durations, our TeleOR system concentrates on identifying and reconstructing only the crucial, dynamic areas. This strategy avoids the unnecessary reconstruction of static parts, thereby simplifying the overall complexity of the reconstruction process.
 
\noindent\textbf{Viewport-Adaptive Transmission:} 
To address the limitations of network bandwidth, particularly in underdeveloped regions, TeleOR incorporates a novel viewport-adaptive transmission strategy. This technique capitalizes on real-time feedback from the client, i.e., viewport information, to selectively transmit and render only those segments of the scene that the user is likely to view, thus ensuring the transmission of high-quality 3D reconstructions in real-time.

Our TeleOR undergoes comprehensive evaluation on 4D-OR dataset, focusing on reconstruction quality and transmission efficiency under various network constraints. The findings reveal that TeleOR achieves high-quality reconstructions and maintains real-time performance even in bandwidth-limited scenarios, highlighting its effectiveness and reliability for tele-intervention applications.

% Our evaluation focuses on quantifying TeleOR's reconstruction accuracy and transmission efficiency, providing empirical evidence of its efficacy in facilitating real-time telemedicine interventions. {\color{red}Furthormore, vis...}

% \textbf{Contributions.} First, TeleOR represents a seminal innovation poised to revolutionize the landscape of telemedicine interventions, transcending the constraints imposed by offline reconstruction methodologies. Second, through meticulous technical design and rigorous validation, TeleOR embodies a paradigm shift towards seamless, real-time engagement within the operating room environment for remote professionals. Third, extensive experiments verify that 

\section{Related Work}
\paragraph{\textbf{Operating Room (OR) Reconstruction.}}
Holistic reconstruction and comprehension of OR marks a pivotal advancement towards the evolution of computer-assisted surgical interventions. 
The integration of sensing technologies within the OR facilitates the identification of individuals, objects, and their interrelations, serving as a foundation for the development of advanced surgical intervention~\cite{gerats2023dynamic,wang2022neural,eck2023real}. 
In this regard, Özsoy et al.\cite{4D_OR} introduced the first public 4D surgical scene dataset, 4D-OR.
% , featuring ten simulated total knee replacement operations captured via six RGB-D sensors in a sophisticated OR simulation facility. 
Gerats et al.\cite{gerats2023dynamic} investigated the application of Neural Radiance Fields (NeRF)~\cite{kato2018neural} for dynamic scene reconstruction within the OR, demonstrating depth-supervised regularization notably enhances image fidelity. 
Additionally, Eck et al.~\cite{eck2023real} employed a marching cubes variant on point clouds for OR surface reconstruction. 
Nevertheless, current approaches have yet to achieve real-time remote surgical intervention, a critical capability for enabling professionals to conduct remote surgical interventions.
% Nevertheless, current initiatives have not realized real-time reconstruction, while which is necessary for facilitating remote interventions by professionals.
% Instead, our \name surpasses these constraints through the innovative use of viewport-dependent, selective reconstruction of the operating room (OR), enabling instantaneous reconstruction. This advancement holds promise for remote transmission and immediate intervention, marking a substantial leap forward in the domain.

% \subsection{Tele-intervention or Tele-consulting or Telemedicine}
\paragraph{\textbf{Virtual Reality (VR) assisted Intervention.}}
Current surgical interventions~\cite{chen2021electrocardio,cannizzaro2022augmented,jud2020applicability,linxweiler2020augmented,zorzal2020laparoscopy,chong2022virtual,doughty2021surgeonassist} typically involve the reconstruction of lesions and organs using VR systems and head-mounted displays, aiding physicians in executing on-site procedures.
Chong et al.~\cite{chong2022virtual} introduced a cutting-edge VR system capable of tracking the laparoscope's pose and reconstructing the 3D surface structure of organs, thereby facilitating laparoscopic surgery. Similarly, Doughty et al.~\cite{doughty2021surgeonassist} developed SurgeonAssist-Net, a framework designed for action-and-workflow-driven virtual surgical assistance. However, a gap remains in real-time remote interventions based on reconstructed OR scene, especially for underdeveloped regions with limited network resources. 
% Tele-intervention plays a crucial role in delivering timely and dependable medical support to remote or underdeveloped regions.

\section{Methodology}
\begin{figure*}
    \centering
    \includegraphics[width=\columnwidth]{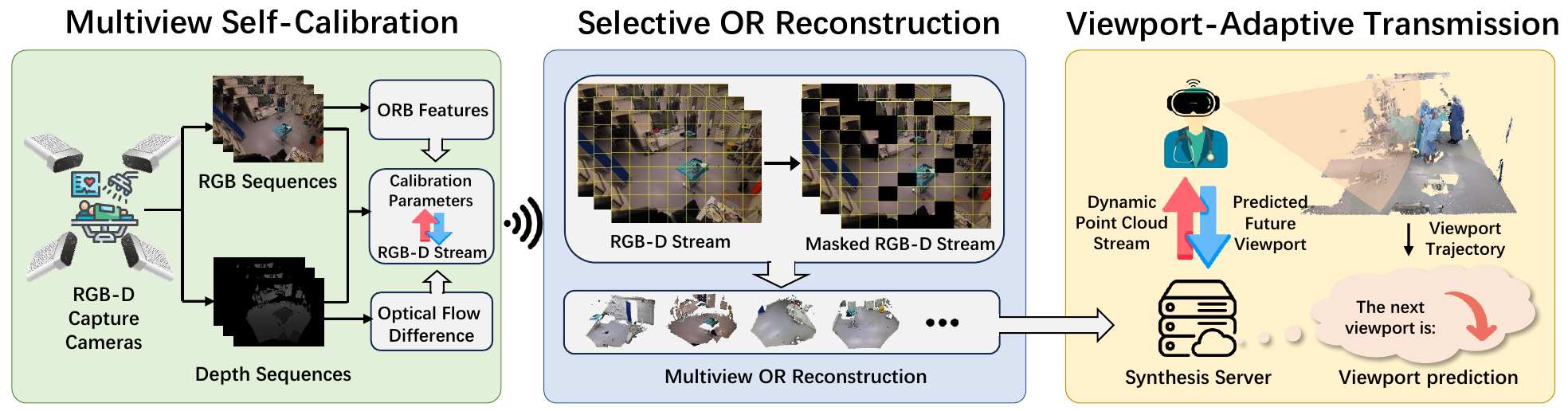}
    \caption{
TeleOR facilitates real-time observation of surgeries through: (1) self-calibration of multiview cameras, (2) selective reconstruction of changing scene segments, and (3) partial remote transmission tailored to the users' predicted viewport for the next time step.}
    \label{fig: system architecture}
    \vskip -15pt
\end{figure*}

In Fig.~\ref{fig: system architecture}, our proposed TeleOR systematically integrates multiview self-calibration, selective OR reconstruction, and viewport-adaptive transmission to provide remote professionals with real-time fully immersive tele-consulting capabilities. 

\subsection{Multiview Self-calibration}
% \noindent\textbf{Mutliview Calibration} is the basis of conducting full-scene OR reconstruction thus providing the professionals with an immersive perception of the remote surgical scene that directly affects the performance of the tele-intervention/tele-consulting. 
Different from traditional indoor reconstruction scenarios, real-time OR reconstruction poses multiple challenges: 
\textbf{(1)} Obstructions of capture devices: 
% Due to obstructions commonly encountered during surgeries, capture devices in OR scene frequently need to be moved, which affects the calibration and synchronization of the capture device array. 
Due to frequent movements of medical staff and equipment during surgeries, capture devices are often obstructed, impacting the calibration and synchronization of the capture device array.
\textbf{(2)} Restriction of pre-recording setups: OR is often equipped with sophisticated instruments and monitors, which prevents the deployment of physical calibration objects exploited by traditional methods. 
To tackle the above challenges, different from traditional calibration methods~\cite{LiveScan3D} that require pre-defined makers priors, we propose a dynamic self-calibration strategy that directly exploits the features extracted in the OR scene as the calibration reference. 

In detail, RGB-D cameras are strategically positioned to ensure overlapping fields of view (FoV), with one camera designated as the anchor. The coordinate system of the anchor camera is set as the origin for the entire coordinate system. Initially, the first frame from each camera is processed to extract ORB features~\cite{karami2017image} for calibration. These features are then matched across adjacent FoV to establish camera positions relative to each other. Subsequently, a transformation matrix is derived to align the reconstructed scene with the reference coordinate system through translational and rotational adjustments.

After the initial calibration, to prevent capture cameras from being obstructed, we mount the cameras on mobile platforms to enable their movement. In this context, preserving calibration accuracy is crucial throughout the surgical procedure due to shifts in the camera's FoV. Therefore, we perform real-time updates of the camera positions. The optical flow method~\cite{liu2013optical} is utilized to track changes in camera pose by estimating the motion of feature points across frames, allowing for effective camera movement tracking. Through continuous analysis of optical flow vectors, we dynamically adjust the camera's calibration parameters, ensuring accurate and synchronized scene capture within the OR system.

\subsection{Selective OR Reconstruction}
%Using optical flow method to detect which part has moved, only update this part 
% Full-scene indoor reconstruction often introduces huge computation consumption and large data size in previous work, which impedes its application in telemedicine intervention that has strict requirements for real-time performance. In this paper, to enable real-time streaming of the generated OR scene with low bandwidth cost, we reuse the reconstructed scene by exploiting an optical-flow-based detection method. 
\textbf{A distinctive characteristic of the OR scene is that many objects remain static for the majority of the surgical process.} Leveraging this observation, we propose to reuse already reconstructed content, focusing updates only on the changing parts of the scene.
To this end, we use an optical-flow-based motion detection method to guide this selective reconstruction procedure. 
% This approach optimizes computational efficiency and ensures the reconstruction process remains both accurate and resource-effective, by selectively updating areas of the scene where significant changes occur.

\begin{figure*}
    \centering
    \includegraphics[width=\linewidth]{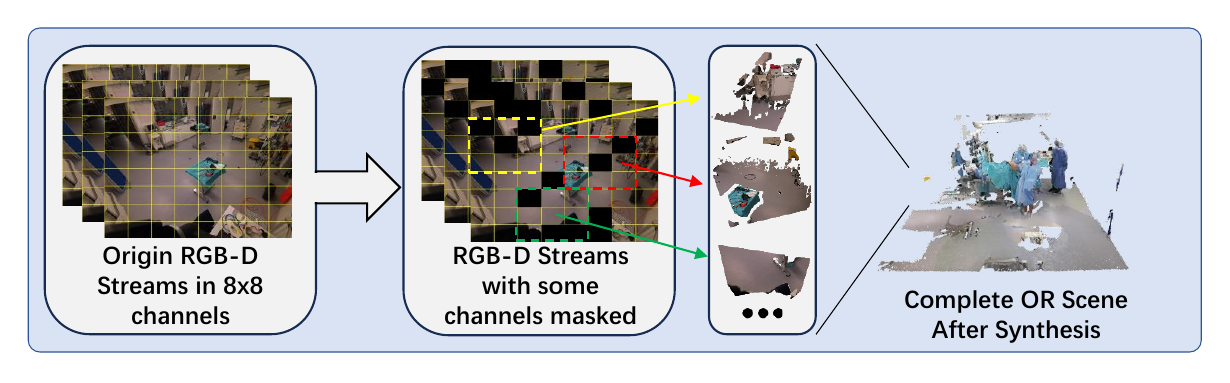}
    \caption{Selective OR Reconstruction Pipeline.}
    \vskip -15pt
    \label{fig: OR_reconstruction}
\end{figure*}

\noindent\paragraph{\textbf{Optical Flow Calculation.}}
% To reduce the computation consumption of OR reconstruction and save transmission bandwidth cost, we design a tile-based selective OR reconstruction algorithm. 
As illustrated in Fig. \ref{fig: OR_reconstruction}, we divide the RGB-D stream from each FoV into N $\times$ N channels, which produces N $\times$ N tiles per frame, named $T_{t}=\left\{  T_{t}^1,T_{t}^2,...,T_{t}^{N^2} \right\}$ for tiles at time \textit{t}. During the reconstruction process of each frame, the inter-frame disparity of each tile is examined, by calculating the cumulative magnitude of the optical flow field difference $D_c$ with the previous tile at the same position. 
To reduce the computational cost of pixel-wise optical flow difference, we exploit sparse optical flow calculation of \texttt{Lucas-Kanade alogrithm}~\cite{yedjour2021optical}. Specifically, we compute the frame-wise optical flow difference based on two assumptions:
\begin{itemize}
    \item \textbf{Brightness Constancy Assumption:} The brightness of a pixel does not change as it moves from one frame to the next. 
    % This is known as the brightness constancy assumption.
    \item \textbf{Small Motion Assumption:} The motion between frames is assumed to be small, and the displacement vector is constant within the local neighborhood.
\end{itemize}
The calculation of optical flow vector ($U_{xy}$, $V_{xy}$) for point (x,y) between two consecutive tiles $T_{t}^n$ and $T_{t-1}^n$, can be attained by solving the following equation:
\begin{equation}
    \begin{bmatrix} U_{xy} \\ V_{xy} \end{bmatrix} = \begin{bmatrix} E_x & E_{xy} \\ E_{xy} & E_y \end{bmatrix}^{-1} \begin{bmatrix} -E_{xt} \\ -E_{yt} \end{bmatrix},
\end{equation}
where
\begin{equation}
    E_x = T_x^2, \quad E_y = T_y^2, \quad E_{xy} = T_x \cdot T_y, \quad E_{xt} = T_x \cdot T_t, \quad E_{yt} = T_y \cdot T_t,
\end{equation}
 \begin{equation}
    T_x(x, y) = \frac{\partial T}{\partial x}\Bigr|_{(x, y)}, \quad T_y(x, y) = \frac{\partial T}{\partial y}\Bigr|_{(x, y)},
\end{equation}
\begin{equation}
    T_t(x, y) = T_{t}(x, y) - T_{t-1}(x, y).
\end{equation}
In this way, we can determine which tile has perceptible visual change given an adjustable threshold according to available computational and transmission bandwidth resources. The calculation of the cumulative magnitude of optical flow difference can be represented as:
\begin{equation}
     D_{c} = \sum_{x=1}^{N_{width}}  \sum_{y=1}^{N_{Height}} M_t(x,y),
\end{equation}
\begin{equation}
    M_t(x,y) = \sqrt{U_{xy}^2+V_{xy}^2}, 
\end{equation}
where $D_{c}$ is the cumulative magnitude of optical flow difference between tile $T_{t}^n$ and $T_{t+1}^n$, and $M_t(x,y)$ is magnitude of the optical flow difference for point (x,y).

\noindent\textbf{Tile-based Selective Reconstruction.}
\label{sec:recon}
% For each tile with $D_c$ over the designated threshold, which is regarded as changing part of the scene. Then, its corresponding position in the RGB-D frame is masked. 
For each tile, if $D_c$ exceeds the set threshold $\theta$, indicating a perceptible visual difference, the corresponding position in the RGB-D frame will be masked. The value of threshold $\theta$ is dynamically determined by the real-time bandwidth.
Then, these partially masked frames are forwarded to the synthesis server for conversion into point clouds. Throughout the point cloud generation process, for the masked tiles within each frame, the corresponding sections of the scene are not regenerated. Instead, the previously generated segments in the preceding frame are retained and reused.

\subsection{Viewport-Adaptive Transmission}
\label{sec:viewport}
To transmit high-quality 3D reconstructions while overcoming limited bandwidth networks, TeleOR capitalizes on real-time feedback from the client, i.e., viewport information, to adaptively load data for transmission. Initially, we gather data on the user's past FoV movements to construct a historical trajectory. Utilizing this data, following~\cite{FoV_prediction}, an efficient LSTM network is employed to predict the user's FoV for the next time step. 

This prediction enables the strategic synthesis of point clouds solely within the predicted FoV, ensuring efficient utilization of bandwidth by transmitting and rendering only the scene segments likely to be in the user's view. 
% In practice, the FoV prediction often introduces some error, which may lead to a cracked frame, to accommodate the inaccurate FoV prediction results, TeleOR employs a larger viewport by expanding the predicted FoV, which covers more areas outside the user's viewing frustum~\cite{viewing_frustum}. Also, to reduce data usages, the point cloud density level is reduced from the center of the predicted FoV. 
% In practice, to ensure accurate predictions, TeleOR adopts a prediction range of 110 degrees, which extends beyond the user's usual viewing frustum~\cite{viewing_frustum}, covering a broader area for enhanced situational awareness. Also, to reduce data usages, the point cloud density level is reduced from the center of the predicted FoV. 
In practice, to ensure accurate predictions and prevent prediction failures, TeleOR employs a prediction range of 120 degrees, expanding beyond the typical user's viewing frustum~\cite{hu2023understanding,viewing_frustum}. This approach not only covers a wider area for improved situational awareness but also strategically reduces the point cloud density from the center of the predicted FoV to minimize data usage.

In parallel, TeleOR considers the current network bandwidth to adjust the dynamic threshold $\theta$ for selective OR scene reconstruction (Sec.~\ref{sec:recon}) and the reconstructed point cloud density for the next transmission. This approach allows for the fine-tuning of data transmission, balancing the quality of the point cloud stream against the imperative of maintaining real-time streaming capabilities of the OR scene. Through this balance of predictive modeling and bandwidth management, TeleOR ensures the delivery of essential, high-quality 3D content under the constraints of varying network conditions.

\section{Experiment}

\begin{figure}[t]
    \centering
    \begin{minipage}[b]{0.48\textwidth}
        \centering
        \includegraphics[width=\linewidth]{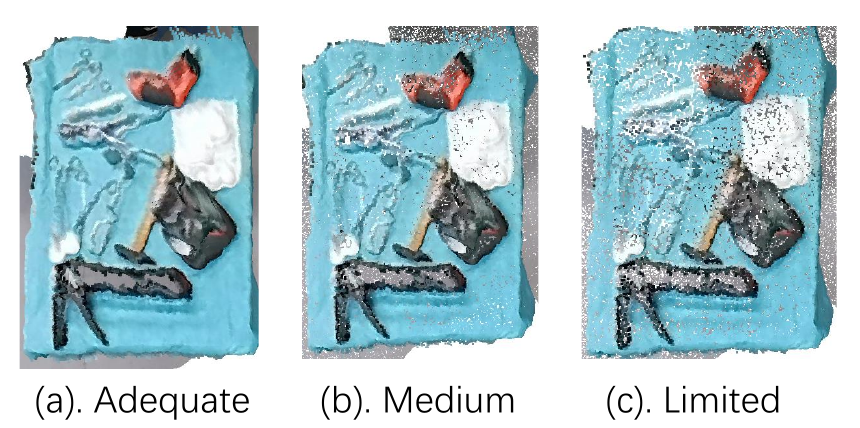}
        \caption{Visualization of reconstruction results of different network conditions.}
        \label{fig:qualitative_results}
    \end{minipage}%
    \hfill%
    % \vspace{-15pt}
    \begin{minipage}[b]{0.48\textwidth}
        \centering
        \renewcommand{\arraystretch}{1.5} % 增加表格行间距
        
        \begin{tabular}{@{}lcc@{}} % @{}去除表格两边的边距
            \toprule
            Net Condition & $R_{reuse}$ & $MSSIM$ \\
            \midrule
            Limited (20Mbps) & 85.8\% & 0.9012 \\
            Medium (50Mbps) & 66.9\% & 0.9615 \\
            Adequate (100Mbps) & 27.3\% & 0.9827 \\
            \bottomrule
        \end{tabular}
        \captionof{table}{Reconstruction quality assessment of different network conditions.}
        % \vspace{10pt}
        \label{table:Reconstruction}
    \end{minipage}
\end{figure}

\subsection{Experimental Setup}
\paragraph{\textbf{Datasets and Evalutaion Metrics.}} To evaluate the reconstruction quality and real-time performance of our proposed TeleOR system, we conduct the evaluation on the public operating room dataset 4D-OR~\cite{4D_OR}, which is composed of 6734 scenes captured by six RGB-D Kinect sensors. The evaluation focuses on two key aspects that have the most significant impact on the tele-intervention performance: reconstruction quality (Sec.~\ref{exp:qua}) and transmission efficiency (Sec.~\ref{exp:eff}).

\noindent \paragraph{\textbf{Implementations.}} We implement the TeleOR on top of Open3D. 
% To avoid ethical questions, the testbed is implemented on the open-source 4D-OR dataset for simulation. 
We deploy the RGB-D sequences of each FoV on six individual edge computing devices (Nvidia Jetson Nano), each connecting to the synthesis server (equipped with Intel i7-12700 CPU, NVIDIA GeForce RTX 3090 GPU, 64G RAM) via a WiFi connection. For the client-side renderer, we built a real-time player on top of Unity and deployed it on a Meta Quest Pro head-mounted display. 
The supplementary materials provide a detailed description of the rendering data format.

\subsection{Reconstruction Quality}
\label{exp:qua}
% Currently, there are no well-defined metrics for reconstructed scene visual quality assessment, thus we propose using SSIM (structural similarity index metric)~\cite{SSIM}, a popular quality metric for 2D visual quality that measures the perceptual quality difference between two videos. 
% To evaluate the visual quality of reconstructed scene,
% we adopt the structural similarity index metric (SSIM) of the 2D rendering of the 3D scene from a predefined set of views. 
% To assess the visual quality of the reconstructed scene, we employ the Structural Similarity Index Metric (SSIM) to analyze the 2D renderings generated from the 3D scene. This evaluation is conducted from a predefined set of viewpoints, which is named as Multiview SSIM (MSSIM), , which indicates the average SSIM across multiple predefined viewports.
% We refer to our metric as Multiview SSIM, which indicates the average SSIM across multiple predefined viewports. 

To accurately assess the visual quality of the reconstructed scene, we use the Structural Similarity Index Metric (SSIM) for analyzing 2D renderings derived from the 3D scene. This analysis is carried out from a series of predefined viewpoints, termed Multiview SSIM (MSSIM). MSSIM represents the average SSIM value obtained across these multiple predefined viewpoints. 
Also, Scene Reuse Ratio ($R_{reuse}$) is adopted to evaluate the effectiveness of utilizing previously reconstructed scene elements in new reconstructions.
We evaluate the reconstruction performance of our selective reconstruction approach under various network constraints, with bandwidth limits set to 20, 50, and 100 Mbps.
% \footnote{It is noteworthy that the typical available bandwidth in hospitals ranges from 100 Mbps to 1 Gbps. 
Therefore, our established bandwidth testing range represents a stringent set of extreme conditions.
For one thing, in Tab.~\ref{table:Reconstruction}, it is observed that at a bandwidth of 20 Mbps, our method achieves a $R_{reuse}$ of 85.8\%, indicating a substantial reuse of segments reconstructed from previous frames. This effectively reduces the complexity of the reconstruction process, ensuring the \textbf{real-time} \textbf{performance} of our approach.
For another, Fig.~\ref{fig:qualitative_results} demonstrates that our TeleOR achieves high-quality reconstructions at 100 Mbps, and maintains acceptable performance even under severe bandwidth restrictions. This significantly ensures the \textbf{visual quality} of real-time reconstruction in bandwidth-constrained environments.

\begin{table}[t]
  \centering
  \caption{Transmission performance under different network conditions.}
  % and ablation study without our key designs.}
  % \adjustbox{max width=\textwidth}{
  \setlength{\tabcolsep}{2mm}{
  \begin{tabular}{lcccccc}
    \toprule
    \multirow{2}{*}{Methods \& Net Condition} & \multicolumn{3}{c}{Frames per Second} & \multicolumn{3}{c}{Latency (ms)} \\
    \cmidrule(lr){2-4} \cmidrule(lr){5-7}
    & Min. & Avg. & Max. & Min. & Avg. & Max. \\
    \midrule
    Limited (20Mbps) & 15.6 & 22.7 & 24.3 & 297 & 302 & 357 \\
    Medium (50Mbps) & 22.5 & 25.2 & 29.9 & 160 & 192 & 269 \\
    Adequate (100Mbps) & 25.9 & 27.7 & 29.9 & 95 & 105 & 136 \\
     \midrule
    w/o. selective recon. (100Mbps) & 3.2 & 5.7 & 6.7 & 132 & 153 & 275 \\
        w/o. viewport adapt. (100Mbps) & 7.5 & 9.2 & 12.7 & 172 & 197 & 267 \\
    \bottomrule
      \end{tabular}}
  \label{tab:transmission}
\end{table}

\subsection{Transmission Efficiency}
\label{exp:eff}
The transmission efficiency is assessed through two metrics: Frames per Second (FPS) and Latency, with FPS reflecting the smoothness of video playback and Latency measuring the delay between an action and its visual feedback on the screen.
Evaluation is conducted on all ten groups of 4D-OR, with each containing 3000 sample frames for simulation, using the same network constraints as Sec.\ref{exp:qua}. 
As shown in Tab.~\ref{tab:transmission}, despite operating under severely constrained bandwidth conditions, TeleOR is still capable of maintaining a frame rate of 22.7 FPS.
% \footnote{For reference, the standard frame rate for movies is 24 FPS, which is considered sufficient to create a smooth sense of motion.}. 

\subsection{Ablation Study}
To explore the effectiveness of TeleOR's key designs, the selective reconstruction (Sec.~\ref{sec:recon}) and viewport adaptive (Sec.~\ref{sec:viewport}) strategy is removed, respectively.
In Tab.~\ref{tab:transmission}, without selective reconstruction or viewport adaptation, a noticeable drop in FPS can be observed even under adequate network resources. 
This observation emphasizes the critical role of TeleOR's core components, especially under poor network conditions.

%QoE evaluation: latency, visual quality(SSIM: reconstructed vs orginal 2D frame), bandwidth consumption, fps, 
\section{Conclusion}
In this paper, we introduced TeleOR, an advanced system designed for enhancing tele-intervention in OR through real-time, high-fidelity scene reconstructions.
By integrating dynamic self-calibration, selective reconstruction, and viewport-adaptive transmission, it addresses issues like camera occlusions, scene complexity, and bandwidth constraints, enhancing remote surgical guidance. 
% Experiments on the 4D-OR dataset indicate that our TeleOR not only shows excellent quality reconstruction accuracy but also higher transmission efficiency even under stringent bandwidth constraints to achieve real-time intervention. 
Experiments on the 4D-OR dataset indicate that our TeleOR not only achieves exceptional reconstruction accuracy but also superior transmission efficiency under strict bandwidth limitations, enabling seamless real-time interventions.

% \vspace{-1em}
% \subsubsection{Acknowledgments.}
% This research was partially supported by National Natural Science Foundation of China under grants No. 62176231, No. 62106218, No. 82202984, No. 92259202 and No. 62132017, Zhejiang Key R\&D Program of China under grant No. 2023C03053.
% \vspace{-1em}
% \subsubsection{Disclosure of Interests.}
% The authors have no competing interests to declare that are relevant to the content of this article.

\bibliographystyle{splncs04}
\bibliography{Paper-0757}
\end{document}